\documentclass[prl,twocolumn,showpacs]{revtex4-1}
\usepackage{graphicx,color}

\begin{document}


\title{Improved tests of Local Position Invariance\\
 using $^{87}$Rb and $^{133}$Cs fountains}


\author{J. Gu\'ena$^1$}
\author{M. Abgrall$^1$}
\author{D. Rovera$^1$}
\author{P. Rosenbusch$^1$}
\author{M.~E. Tobar$^2$}
\author{Ph. Laurent$^1$}
\author{A. Clairon$^1$}
\author{S. Bize$^1$}
\affiliation{$^1$LNE-SYRTE, Observatoire de Paris, CNRS, UPMC, 75014 Paris, France}
\affiliation{$^2$School of Physics, University of Western Australia, Crawley, Australia}


\date{\today}


\begin{abstract}

We report tests of local position invariance based on measurements of the ratio of the ground state hyperfine frequencies of $^{133}$Cs and $^{87}$Rb in laser-cooled atomic fountain clocks. Measurements extending over 14~years set a stringent limit to a possible variation with time of this ratio: $d \ln(\nu_{\mathrm{Rb}}/\nu_{\mathrm{Cs}})/dt=(-1.39\pm 0.91 )\times 10^{-16}$~yr$^{-1}$. This improves by a factor of 7.7 over our previous report [H.~Marion \textit{et al.}, Phys.~Rev.~Lett.~\textbf{90}, 150801 (2003)]. Our measurements also set the first limit to a fractional variation of the Rb/Cs frequency ratio with gravitational potential at the level of $c^2 d \ln(\nu_{\mathrm{Rb}}/\nu_{\mathrm{Cs}})/dU=(0.11 \pm 1.04)\times10^{-6}$, providing a new stringent differential redshift test. The above limits equivalently apply to the fractional variation of the quantity
$\alpha^{-0.49}\times(g_{\mathrm{Rb}}/g_{\mathrm{Cs}})$, which involves the fine-structure constant $\alpha$ and the ratio of the nuclear $g$-factors of the two alkalis. The link with variations of the light quark mass is also presented together with a global analysis combining with other available highly accurate clock comparisons.
\end{abstract}


\maketitle




Einstein's equivalence principle is one of the founding principle of general relativity.
Many experiments have been dedicated to testing the validity of this principle \cite{Will2006},
several of them searching for variations of fundamental constants, either on cosmological timescales using astronomical and geochemical data or in our present epoch, by exploiting highly accurate atomic clocks. Such variations would violate local position invariance (LPI), one of the three components of Einstein's equivalence principle. The possibility that dimensionless fundamental constants might change in time or space is allowed or predicted by alternative theories aimed at unifying gravitation with the other fundamental interactions, hence the strong interest in this search, which could reveal physics beyond general relativity and the Standard Model of particle physics. There has been a claim of a variation of the fine-structure constant $\alpha$ over cosmological time scales from the analysis of Quasar absorption spectra \cite{Webb2001}. This was followed by other conflicting observations \cite{Srianand2004} and by several controversies (see, for instance, Ref.~\cite{Uzan2011}). Recently, it was suggested that conflicting observations could be reconciled if one assumes that $\alpha$ varies in space rather than in time \cite{Webb2011}. Atomic clocks offer the possibility to search for variation of constants at the present epoch in laboratory-based experiments whose interpretations are fully independent of any cosmological model \cite{Prestage1995,Marion2003,Fischer2004,Fortier2007,Cingoz2007,Ferrell2007,Ashby2007,Blatt2008,Rosenband2008,Peik2010}.

In this letter, we present new LPI tests obtained by comparing the ground state hyperfine frequencies of $^{133}$Cs and $^{87}$Rb atoms over a period of $\sim 14$~years. Our measurements give the second most stringent limit to date to a possible variation with time of the ratio of two atomic frequencies, and the most stringent for two hyperfine frequencies. They also provide the first limit to a variation of the hyperfine frequency ratio $\nu_{\mathrm{Rb}}/\nu_{\mathrm{Cs}}$ with gravitational potential. The link with fundamental constants of the Standard Model is made with atomic and nuclear structure calculations \cite{Flambaum2006,Dinh2009}. Combining our measurements with other clock experiments, we set updated limits to variations of the fine-structure constant $\alpha$, of the electron-to-proton mass ratio $\mu=m_{e}/m_{p}$ and of the ratio of the light quark mass to the Quantum Chromodynamics mass scale  $m_{q}/\Lambda_{\mathrm{QCD}}$.

\begin{figure}
 \includegraphics*[width=8.5cm]{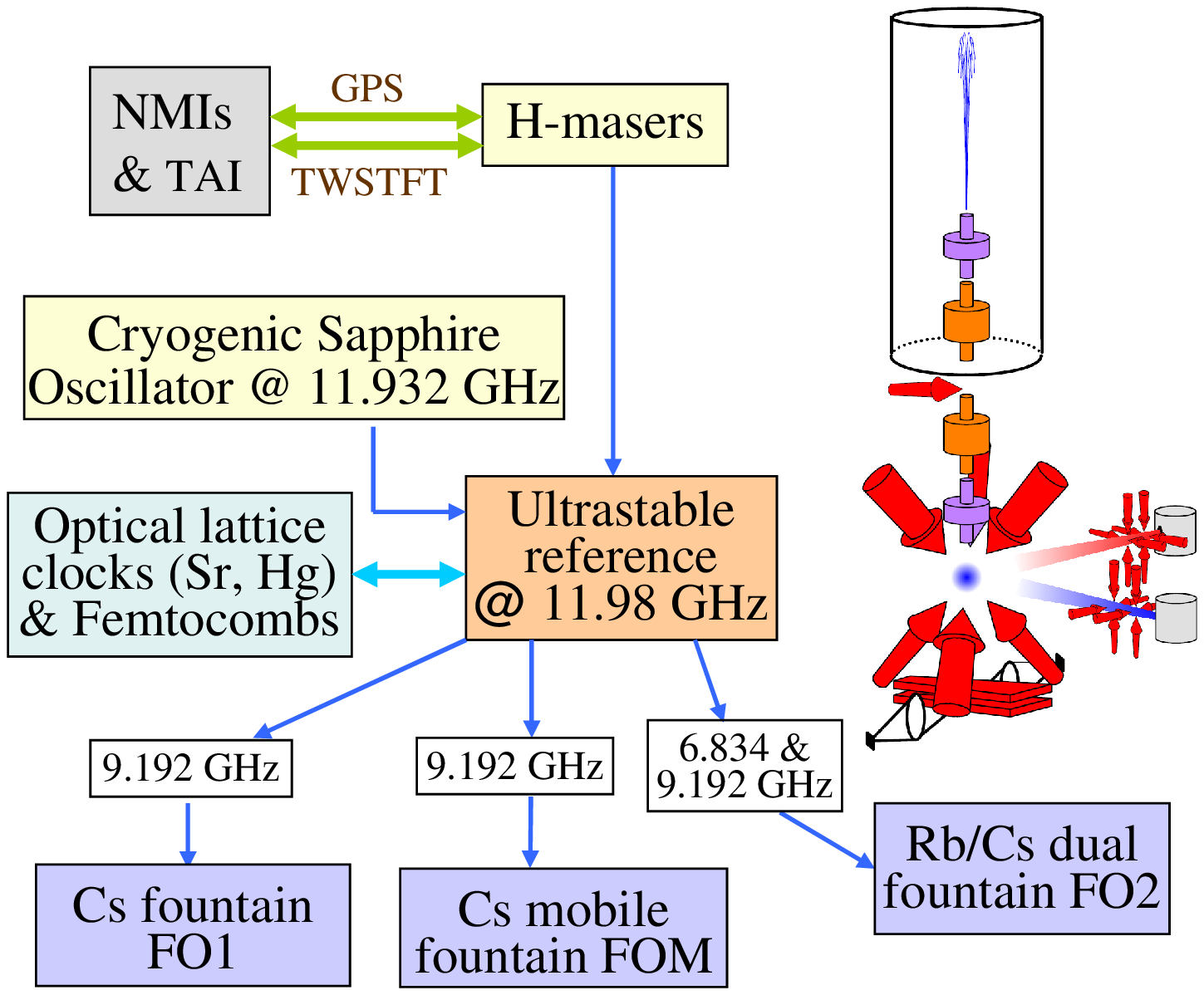} 
 \caption{LNE-SYRTE atomic clock ensemble. The block diagram shows the 3 fountains FO1, FO2 and FOM connected to the common ultra-low noise cryogenic sapphire oscillator. The FO2 fountain, schematized on the right, can operate with $^{87}$Rb and $^{133}$Cs simultaneously, enabling the present hyperfine frequency comparisons. The optical clocks, the H-masers, the sapphire oscillator, as well as the links to other remote laboratories and to international timescales (TAI is Temps Atomique International) via satellite time transfers (GPS, TWSTFT) provide for further tests of Einstein's equivalence principle \cite{{Fischer2004},{Ashby2007}, {Blatt2008},{Tobar2010},{Wolf2006}}.
\label{FountainEnsemble}}
 \end{figure}

Our measurements exploit the LNE-SYRTE atomic fountain ensemble schematized in Fig. \ref{FountainEnsemble},
 which consists of three $^{133}$Cs Primary Frequency Standards FO1, FO2-Cs and FOM, sharing
 a common ultra-low noise cryogenic oscillator. Among them, FO2 is a dual fountain that can also operate with $^{87}$Rb. The three Cs clocks operate on the hyperfine clock transition at 9.2~GHz and the Rb clock on the corresponding 6.8~GHz transition. A detailed description of the latest developments of this fountain ensemble is given in Ref.~\cite{Guena2012}. Notable advances relevant to the present work were the simultaneous operation with Rb and Cs in FO2 \cite{Guena2010} and large improvements in reliable, unattended operation for all fountains, allowing for quasicontinuous Rb/Cs comparisons for several months. Also, fountain accuracy improved along the years, notably with the several recent studies of systematic shifts \cite{{Rosenbusch2007},{Guena2010},{Guena2011}, {Li2011},{Guena2012}}. In FO2, Rb and Cs atoms are simultaneously laser-cooled, launched, state-selected, and probed with the Ramsey interrogation method, and finally selectively detected by time-resolved laser-induced fluorescence, in the same vacuum chamber \cite{Guena2010}, as shown in Fig.~\ref{FountainEnsemble}. Rb and Cs are launched at slightly different velocities to separate the two clouds during interrogation and thereby avoid interspecies collisions. Typical accuracies over the recent years relevant to this work are $4-5 \times 10^{-16}$ for FO1, FO2-Cs and FO2-Rb, and $7-8 \times 10^{-16}$ for the transportable fountain FOM. The detailed systematic uncertainty budgets for the four clocks are available in Table III of Ref.~\cite{Guena2012}. The four clocks measure the frequency of the common local oscillator (Fig. \ref{FountainEnsemble}), with cycle times ranging from 1.1~s to 1.6~s. The data from each clock are corrected for all systematic shifts (as described in Ref.~\cite{Guena2012}, Sect. IV.A)  and averaged over synchronous intervals of 864~s. Next, for each available Cs fountain, the Rb/Cs frequency ratio is determined over these synchronous intervals, which removes the frequency of the common local oscillator. Typically, the fractional frequency instability of a Rb/Cs comparison is limited by quantum projection noise, and ranges from $8 \times 10^{-14}$ to $1.2 \times 10^{-13}$ at 1~s.  A fractional resolution of $2 \times 10^{-16}$ for the $\nu_{\mathrm{Rb}}/\nu_{\mathrm{Cs}}$ ratio is reached in a few days of averaging time, after which the overall uncertainty of the comparison becomes limited by systematic uncertainties.

 \begin{figure}
 \includegraphics*[width=8.5cm]{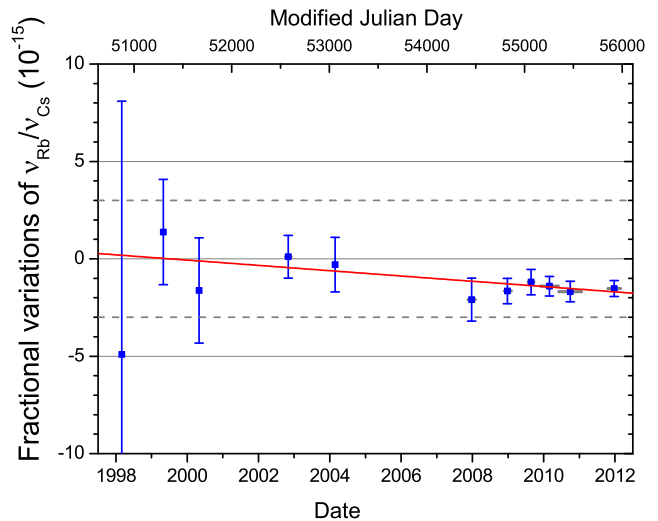} 
\caption{Temporal record of fractional variations of the $\nu_{\mathrm{Rb}}/\nu_{\mathrm{Cs}}$ hyperfine frequency ratio. The error bars are the total 1~$\sigma$ uncertainties, dominated by the systematic uncertainties. The horizontal bars show the duration of each comparison. The solid line is the weighted fit to a line with inverse quadratic weighting. The origin of the vertical axis corresponds to the $^{87}$Rb secondary representation of the SI second recommended by BIPM, with a recommended uncertainty $3 \times 10^{-15}$ (dashed lines) \cite{CCTF2004}. \label{FitLin}}
 \end{figure}

Figure \ref{FitLin} presents the temporal record of the Rb/Cs comparisons since the first operation of FO2-Rb in 1998. Up to 2008 (the first six points), FO2 was operated either with $^{87}$Rb only or alternately with $^{87}$Rb and $^{133}$Cs atoms, and the Cs reference was FO1 or FOM, or both. Since the end of 2008, FO2 has been operating with Rb and Cs simultaneously and the dominant Cs reference is FO2-Cs. Each point in Fig.~\ref{FitLin} represents the average over a duration ranging from a few weeks to several months, the averaging periods being chosen based on significant modifications on either fountain. When two or three Cs references are available, we compute a weighted average taking into account the total uncertainty and the amount of measurement time for each Cs/Rb pair. Error bars in Fig.~\ref{FitLin} are the overall one standard deviation (1~$\sigma$) uncertainties, which are dominated by systematic uncertainties.

The solid red line in Fig. \ref{FitLin} is the result of the weighted linear least-squares fit of a straight line to the data with inverse quadratic weighting, \textit{i.e.} weights inversely proportional to the square of the error bars:
\begin{equation}\label{drift}
   \frac{d}{dt}\ln\left(\frac{\nu_{\mathrm{Rb}}}{\nu_{\mathrm{Cs}}}\right)= (-1.36\pm 0.91)\times 10^{-16}\mathrm{yr}^{-1}.
\end{equation}
The yearly drift deviates from zero by 1.5~$\sigma$ which is not statistically significant.
The uncertainty of this result improves by a factor 7.7 over our 2003 report \cite{Marion2003}.
It is one of the most constraining accurate clock comparisons over long timescales (see Table~\ref{Experiments}).

Hyperfine splitting (hfs) energies scaled to the Rydberg energy ($R_\infty h c$) depend on the fine-structure constant $\alpha$ and on the nuclear $g$-factor. Consequently, the $\nu_{\mathrm{Rb}}/\nu_{\mathrm{Cs}}$ hyperfine frequency ratio is sensitive to variations of $\alpha$ and of the nuclear $g$-factors of the two atoms \cite{Prestage1995}. From the recent atomic structure calculations of the $\alpha$-dependent relativistic effects \cite{Flambaum2006}, we get
$d\ln(\nu_{\mathrm{Rb}}/\nu_{\mathrm{Cs}})$= $d\ln[\alpha^{-0.49}\times(g_{\mathrm{Rb}}/g_{\mathrm{Cs}})]$. Thus, Eq.~\ref{drift} yields
\begin{equation}\label{driftb}
   \frac{d}{dt}\ln\left(\alpha^{-0.49}\times \frac{g_{\mathrm{Rb}}}{g_{\mathrm{Cs}}}\right)= (-1.36\pm 0.91)\times 10^{-16}\mathrm{yr}^{-1}.
\end{equation}
The link between the $g$-factors and the fundamental parameters of the Standard Model was achieved in Ref.~\cite{Dinh2009}, where it is shown that a variation in $g$-factors can be related to a variation of the light quark mass $m_{q}$ scaled to the Quantum Chromodynamics mass scale $\Lambda_{\mathrm{QCD}}$. The relation for the Rb and Cs $g$-factors yields
\begin{equation}\label{driftc}
   \frac{d}{dt}\ln(\alpha^{-0.49}(\frac{m_q}{\Lambda_{\mathrm{QCD}}})^{-0.021})=(-1.36\pm 0.91)\times 10^{-16}\mathrm{yr}^{-1},
\end{equation}
showing that hyperfine transitions can test both the electroweak ($\alpha$) and the strong ($m_{q}/\Lambda_{\mathrm{QCD}}$) interactions, although without distinguishing between the two contributions.
\begin{figure}
\includegraphics*[width=8.5cm]{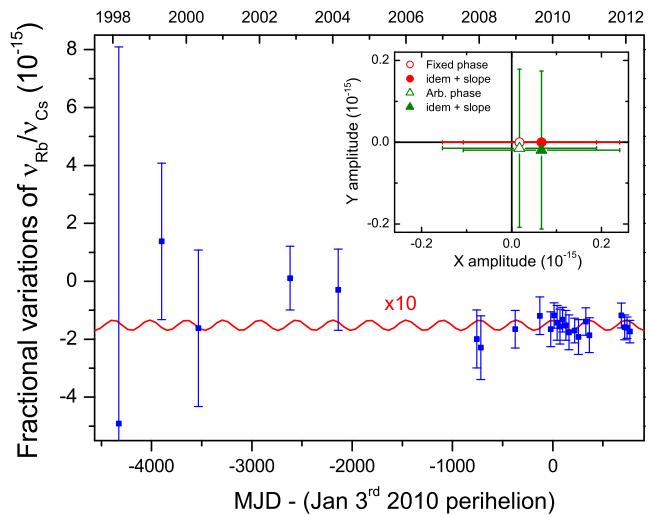} 
 \caption{Search for a modulation of the $\nu_{\mathrm{Rb}}/\nu_{\mathrm{Cs}}$ hyperfine frequency ratio synchronous with the annual change of the solar gravitational potential on Earth. The red curve is the fitted modulation magnified by $10$ for visibility.
 The inset shows the fitted amplitudes in phase (X) and in quadrature (Y) with the potential modulation. Fits with fixed phase (red dots) or arbitrary phase (green triangles), with (filled symbols) and without (unfilled symbols) accounting for a slope. \label{FitModul}} 
 \end{figure}

Next, we search for a possible coupling of the $\nu_{\mathrm{Rb}}/\nu_{\mathrm{Cs}}$ frequency ratio to gravity. For that purpose, we group our Rb/Cs data over shorter intervals of typically one month, as shown in Fig.~\ref{FitModul},
and we look for variations correlated with the annual change of the Sun gravitational potential on Earth, $\Delta U(t)\approx \Delta U \cos[\Omega_\oplus (t-t_{p})]$, where $\Omega_\oplus$ is the Earth's orbital angular frequency, $t_{p}$ a perihelion date, and $\Delta U \approx G M_{\odot} \epsilon/a$,
to first order in the Earth's orbit eccentricity $\epsilon=0.0167$. $G=6.674 \times 10^{-11} $m$^{3}$kg$^{-1}$s$^{-2}$ is the gravitational constant; $M_{\odot}=1.98\times 10^{30}$kg the solar mass; and $a = 1$~AU the orbit semi-axis, yielding $\Delta U/c^{2}\simeq 1.65\times 10^{-10}$, where $c$ is the speed of light. We use the sign convention recommended by the International Astronomical Union \cite{Soffel2003}; i.e., $U(r)\approx +G M_{\odot}/r$ is positive and maximum at perihelion.
The fit function to the data in Fig.~\ref{FitModul} is $\cos[\Omega_\oplus (t-t_{p})]$, the modulation amplitude being a free parameter. For $t_{p}$, we take the Jan 3$^{rd}$, 2010, perihelion: MJD=55199. The fitted amplitude is $(0.18\pm 1.71)\times 10^{-16}$ for the fractional modulation of $\nu_{\mathrm{Rb}}/\nu_{\mathrm{Cs}}$ ratio, which yields
\begin{equation}\label{RbCsdU}
    c^{2}\frac{d}{dU}\ln\left(\frac{\nu_{\mathrm{Rb}}}{\nu_{\mathrm{Cs}}}\right)=(0.11\pm 1.04)\times 10^{-6}.
\end{equation}
This result can be interpreted as a differential redshift experiment, testing whether the gravitational redshift depends on the clock composition. The degree of LPI violation is usually parameterized by a composition-dependent coefficient $\beta$ in the gravitational redshift \cite{Will2006}
$d\nu/\nu=(1+\beta)dU/c^{2}$.
Our measurement yields
\begin{equation}\label{BetaRbCs}
   \beta(^{87}\mathrm{Rb})- \beta(^{133}\mathrm{Cs})=(0.11 \pm 1.04)\times 10^{-6}.
\end{equation}
The previous most sensitive test of this type is based on comparisons between H-masers and Cs fountains \cite{Ashby2007}. Our result is 1.4 times more stringent. Further, it introduces a new atomic transition ($^{87}$Rb hfs).
Finally, it relies on clocks with high accuracy for both Rb and Cs, while H-masers are known to be influenced by ``poorly understood and unmodeled factors" \cite{Ashby2007}.

A violation of LPI in this test could occur via a coupling of fundamental constants to gravity, as predicted by some unification theories.
Using the previously mentioned sensitivity of the $\nu_{\mathrm{Rb}}/\nu_{\mathrm{Cs}}$ ratio to variations of $\alpha$ and of the nuclear $g$-factors, we obtain $c^{2}d \ln(\alpha^{-0.49}\times g_{\mathrm{Rb}}/g_{\mathrm{Cs}})$/$dU$= $(0.11\pm1.04)\times 10^{-6}$, where again the variation of $g_{\mathrm{Rb}}/g_{\mathrm{Cs}}$ can also be presented as a variation of $(m_{q}/\Lambda_{\mathrm{QCD}})^{-0.021}$.

Our experiment can be exploited to bound a possible variation of constants in space. In the Barycentric Celestial Reference Frame, the Earth approximately follows a circular orbit of radius $a=1$~AU at the angular frequency $\Omega_\oplus$. To search for a variation of $\nu_{\mathrm{Rb}}/\nu_{\mathrm{Cs}}$ correlated with this motion, we now fit the data of Fig.~\ref{FitModul} with $\cos(\Omega_\oplus t + \varphi)$ with an arbitrary phase $\varphi$. We find a modulation amplitude $\Delta\ln(\nu_{\mathrm{Rb}}/\nu_{\mathrm{Cs}})= (0.23 \pm 1.80)\times 10^{-16}$. The inset of Fig.~\ref{FitModul} shows the result of this fit in the form of in-phase (X) and in-quadrature (Y) components with respect to the perihelion. From this analysis, we find that $d\ln(\nu_{\mathrm{Rb}}/\nu_{\mathrm{Cs}})/dr \equiv d\ln[\alpha^{-0.49}\times(g_{\mathrm{Rb}}/g_{\mathrm{Cs}})]/dr=(0.23 \pm 1.80)\times 10^{-16}$~AU$^{-1}$ or $(0.15\pm 1.20)\times 10^{-27}$~m$^{-1}$. Another, probably more relevant, approach is to consider the motion of our experiment with respect of the rest frame of the cosmic microwave background, as suggested in \cite{Berengut2012}. This motion is largely dominated by the motion of the solar system at a velocity of $369\pm 0.9$~km.s$^{-1}$ in the direction (168$^\circ$, -7$^\circ$) \cite{Hinshaw2009}. In the cosmic microwave background rest frame, the result of Eq.~\ref{drift} leads to $d\ln(\nu_{\mathrm{Rb}}/\nu_{\mathrm{Cs}})/dR \equiv d\ln(\alpha^{-0.49}\times g_{\mathrm{Rb}}/g_{\mathrm{Cs}})/dR=(-1.17\pm 0.78)\times 10^{-29}$~m$^{-1}$, which is still approximately two orders of magnitude short of resolving the spatial variation suggested in Refs.~\cite{Webb2011,Berengut2012}.

\begin{table*}
\caption{ Sensitivity coefficients k$_{\alpha}$, k$_{\mu}$, k$_{q}$ of atomic transition frequencies used in current atomic clocks to a variation of $\alpha$ \cite{{Dzuba2008},{Flambaum2009}},
of $\mu=m_{e}/m_{p}$ and of $m_{q}/\Lambda_{\mathrm{QCD}}$ \cite{{Flambaum2006},{Dinh2009}}.
These transitions are hyperfine transitions for $^{1}\mathrm{H}_\mathrm{hfs}$, $^{87}\mathrm{Rb}$, $^{133}\mathrm{Cs}$, and optical transitions for $^{1}\mathrm{H(1S-2S)}$ and all others excepted Dy. For Dy, the rf transition between two closely degenerated electronic levels of opposite parity is used in the two 162 and 163 isotopes \cite{Dzuba1999,Cingoz2007,Ferrell2007}.
\label{SensitivityCoeff}}
\begin{tabular}{p{10mm} cccccccccc}
\hline \hline
&$^{87}\mathrm{Rb}$&$^{133}\mathrm{Cs}$& $^{1}\mathrm{H}_\mathrm{hfs}$&$^{1}\mathrm{H(1S-2S)}$&$^{171}\mathrm{Yb}^{+}$& $^{199}\mathrm{Hg}^{+}$&$^{87}\mathrm{Sr}$&$(^{162}\mathrm{Dy}$--$^{163}\mathrm{Dy})$& $^{27}\mathrm{Al}^{+}$\\
\hline
$k_{\alpha}$&2.34 & 2.83 & 2.0 &$\sim$0 & 1.0 & -2.94 & 0.06 & 1.72$\times 10^{7}$& 0.008 \\
$k_{\mu}$&1 & 1 & 1  & 0 & 0 & 0 & 0 & 0&0\\
$k_{q}$&-0.019&0.002 & -0.100 & 0 & 0 & 0 & 0 & 0 & 0\\
\hline
\end{tabular}\\
\end{table*}
\begin{table*}
\caption{Results of atomic frequency ratio measurements used to search for variations of fundamental constants. Second column: sensitivity coefficients deduced from Table \ref{SensitivityCoeff} for each atomic frequency ratio X. Third  column: variation with time in yr$^{-1}$. Last column: variation with the gravitational potential. We give the measured fractional modulation of the frequency ratio X scaled, to solar gravitational potential modulation $\Delta U/c^{2} \approx G M_{\odot} \epsilon/(ac^{2})\approx 1.65\times 10^{-10}$. \label{Experiments}}
\begin{tabular}{p{28mm}|ccc|cc|cc}
\hline \hline
 Frequency ratio $X$ & $k_{\alpha}$ &$k_{\mu}$ &$k_{q}$&$d\ln(\mathrm{X})/dt~~~(\mathrm{yr}^{-1})$&Ref.~ &$c^{2}d\ln(X)/dU$ &Ref.\\
\hline
$\mathrm{Rb}/\mathrm{Cs}$&-0.49 & 0 & -0.021 &$(-1.36\pm 0.91)\times 10^{-16}$&This work &$(0.11\pm 1.04)\times 10^{-6}$&This work\\
$\mathrm{H}_\mathrm{hfs}/\mathrm{Cs}$& -0.83 & 0& -0.102 &   --  && $|0.1\pm 1.40|\times 10^{-6}$&\cite{Ashby2007}\\
$\mathrm{H(1S-2S)}/\mathrm{Cs}$& -2.83 & -1& -0.002&$(-32\pm 63)\times 10^{-16}$&\cite{Fischer2004}& --&\\
$\mathrm{Yb}^{+}/\mathrm{Cs}$& -1.83 & -1& -0.002 &$(-4.9\pm 4.1)\times 10^{-16}$&\cite{Peik2010}& --&\\
$\mathrm{Hg}^{+}/\mathrm{Cs}$&-5.77 & -1& -0.002 &$(3.7\pm 3.9)\times 10^{-16}$&\cite{Fortier2007}&$(2.0\pm 3.5)\times 10^{-6}$&\cite{Fortier2007}\\
$\mathrm{Sr}/\mathrm{Cs}$&-2.77 & -1&-0.002&$(-10\pm 18)\times 10^{-16}$&\cite{Blatt2008}&$(-11.5\pm 18.2)\times 10^{-6}$&\cite{Blatt2008} \\
$(^{162}\mathrm{Dy}$--$^{163}\mathrm{Dy})/\mathrm{Cs}$&1.72$\times 10^{7}$ & -1& -0.002&$(-4.0\pm 4.1)\times 10^{-8}$&\cite{Cingoz2007}&($134\pm 104$)&\cite{Ferrell2007} \\
$\mathrm{Al}^{+}/\mathrm{Hg}^{+}$&2.95 & 0& 0 &$(-0.53\pm 0.79)\times 10^{-16}$&\cite{Rosenband2008}&--&\\
\hline
\end{tabular}
\end{table*}

A possible fractional variation of any atomic frequency X can be related to variations of three dimensionless constants, $\alpha$, $\mu=m_{e}/m_{p}$, and $m_{q}/\Lambda_{\mathrm{QCD}}$. We can write
$d\ln(\mathrm{X}) = k_{\alpha} d\ln(\alpha) + k_{\mu} d\ln(\mu)+k_{q} d\ln(m_{q}/\Lambda_{\mathrm{QCD}})$, where the $k$'s represent the sensitivity coefficients of the specific transition X. This offers the possibility to set independent limits for the three constants using multiple atomic transitions. Next, we provide such a global analysis combining our measurements with other reported accurate clock comparisons. Table \ref{SensitivityCoeff} lists the sensitivity coefficients that we use in this analysis and which are taken from the most recent atomic and nuclear structure calculations \cite{{Dzuba2008},{Flambaum2009},{Flambaum2006},{Dinh2009}}. Sensitivity to $\mu$ and $m_{q}/\Lambda_{\mathrm{QCD}}$ comes from the nuclear magnetic moment involved in hyperfine transitions. Both optical and hyperfine transitions have dependence in the fine structure constant $\alpha$ via relativistic corrections.
Table \ref{Experiments} gives the corresponding sensitivities for clock comparisons of relevance here. Dependence in $\mu$
cancels out in hyperfine-to-hyperfine comparisons. Optical-to-optical comparisons are sensitive to $\alpha$ alone, as exemplified here by the $\mathrm{Al}^{+}/\mathrm{Hg}^{+}$ comparison.

To set independent limits to time variations of constants, we perform a weighted least-squares fit to all experimental results listed in Table \ref{Experiments} (3$^{\mathrm{rd}}$ column), including our result of Eq. \ref{drift}. This fit yields independent constraints for the three constants $\alpha$, $\mu$ and $m_{q}/\Lambda_{\mathrm{QCD}}$ as reported in the first row of Table \ref{globalfit}. The constraint relative to $\alpha$ is mainly determined by the $\mathrm{Al}^{+}/\mathrm{Hg}^{+}$ comparison.
In this fit, only the Rb/Cs comparison disentangles $\mu$ and $m_q/\Lambda_{\mathrm{QCD}}$. It is therefore essential to constrain $m_q/\Lambda_{\mathrm{QCD}}$. This stems from the fact that optical frequency measurements are all performed against Primary Frequency Standards, i.e. against the Cs hyperfine frequency.
We note that the constraint for $\alpha$ is slightly less stringent than in Ref.~\cite{Rosenband2008} because we are using the more recent and reduced sensitivity coefficient of Ref.~\cite{Flambaum2009}.

Similarly, we perform a global analysis for the variation with gravitational potential exploiting the available comparisons (Table \ref{Experiments}, last column). We could find that for the Sr/Cs comparison, the modulation amplitude of the gravitational potential was overestimated by a factor of 2 relative to the frequency modulation in Ref.~\cite{Blatt2008}. The table gives the corrected, 2 times less stringent constraint for this comparison. Also, we have checked the consistency between the conventions that we have chosen (sign of U, phase of the modulation) and those in Refs.~\cite{Ashby2007,Blatt2008,Fortier2007,Ferrell2007}.
The least-squares fit to these results yields independent constraints for the three couplings to gravity, reported in the second row of Table \ref{globalfit}.

Finally, we note that an alternative approach consists in using $\alpha$, $m_{e}/\Lambda_{\mathrm{QCD}}$, and $m_{q}/\Lambda_{\mathrm{QCD}}$ instead of $\alpha$, $\mu$, and $m_{q}/\Lambda_{\mathrm{QCD}}$ as parameters of the Standard Model. In this other approach, the sensitivity coefficients of Table \ref{SensitivityCoeff} become $k'_\alpha=k_\alpha$, $k'_{e}$= $k_{\mu}$ and $k'_{q}$ = $k_{q}$ - 0.048 \cite{Flambaum2006}. The final results of this second analysis is identical for $\alpha$ and $m_{q}/\Lambda_{\mathrm{QCD}}$, and we find that $d \ln[m_{e}/\Lambda_{\mathrm{QCD}}]/dt=(4.9\pm 3.7)\times 10^{-16}\mathrm{yr}^{-1}$ and $c^{2}d \ln[m_{e}/\Lambda_{\mathrm{QCD}}]/dU=(-4\pm 17)\times 10^{-6}$.

\begin{table}
\caption{Results of the global analysis of the atomic clock comparisons given in Table \ref{Experiments}: constraints on temporal variations and couplings to gravitational potential for the three fundamental constants: $\alpha$, $\mu=m_{e}/m_{p}$, and $m_{q}/\Lambda_{\mathrm{QCD}}$.
\label{globalfit}}
\begin{tabular}{p{28mm} ccc}
\hline \hline
 & $\ln(\alpha)$ &$\ln(\mu$)&$\ln(m_{q}/\Lambda_{\mathrm{QCD}})$ \\
\hline
$d/dt~ (\times 10^{-16}\mathrm{yr}^{-1})$& $-0.25\pm 0.26$ &$1.5\pm 3.0$& $71\pm 44$\\
$c^{2}d/dU ~(\times 10^{-6})$ & $0.5 \pm 2.7$ & $-4 \pm 15$ & $-5\pm 28$ \\
\hline
\end{tabular}
\end{table}

We have reported highly sensitive tests of LPI using $^{87}$Rb and $^{133}$Cs atomic fountain clocks. Exploiting also other available clock comparisons, we set stringent constraints to possible variations of $\alpha$, $\mu=m_{e}/m_{p}$ and $m_{q}/\Lambda_{\mathrm{QCD}}$ with time and gravitational potential. The rapid developments seen in atomic clocks, especially in the optical domain, as well as the dramatic improvement in long distance clock comparisons allowed by coherent optical fiber links \cite{Lopez2010,Predehl2012} or the ACES space mission \cite{Cacciapuoti2007} will largely diversify and enhance these tests.

\begin{acknowledgments}
SYst\`emes de R\'ef\'erence Temps-Espace (SYRTE) is UMR 8630 between CNRS, UPMC and Observatoire de Paris. This work is largely funded by LNE. We acknowledge the contribution and support from SYRTE's technical services. We thank S. Blatt for his open and useful discussion of the factor 2 in Ref.~\cite {Blatt2008}.
\end{acknowledgments}

\end{document}